\documentclass[reprint,superscriptaddress,amsmath,amssymb, aps, prapplied]{revtex4-2}

\usepackage{graphicx}
\usepackage{hyperref}
\usepackage{booktabs}
\usepackage{siunitx}[=v2]
\usepackage[utf8]{inputenc}

\begin{document}

\title{Atom-number enhancement by shielding atoms from losses in strontium magneto-optical traps}

\author{Jonatan H\"oschele}%
\affiliation{ICFO - Institut de Ciencies Fotoniques, The Barcelona Institute of Science and Technology, 08860 Castelldefels (Barcelona), Spain}%
\author{Sandra Buob}%
\affiliation{ICFO - Institut de Ciencies Fotoniques, The Barcelona Institute of Science and Technology, 08860 Castelldefels (Barcelona), Spain}%
\author{Antonio Rubio-Abadal}%
\affiliation{ICFO - Institut de Ciencies Fotoniques, The Barcelona Institute of Science and Technology, 08860 Castelldefels (Barcelona), Spain}%
\author{\\Vasiliy Makhalov}%
 \email{vasiliy.makhalov@icfo.eu}%
\affiliation{ICFO - Institut de Ciencies Fotoniques, The Barcelona Institute of Science and Technology, 08860 Castelldefels (Barcelona), Spain}%
\author{Leticia Tarruell}%
\affiliation{ICFO - Institut de Ciencies Fotoniques, The Barcelona Institute of Science and Technology, 08860 Castelldefels (Barcelona), Spain}%
\affiliation{ICREA, Pg. Llu\'{i}s Companys 23, 08010 Barcelona, Spain}

\date{\today}

\begin{abstract}
We present a scheme to enhance the atom number in magneto-optical traps of strontium atoms operating on the 461\,nm transition. This scheme consists of resonantly driving the \textsuperscript{1}S\textsubscript{0}$\to$\textsuperscript{3}P\textsubscript{1} intercombination line at 689\,nm, which continuously populates a short-lived reservoir state and, as expected from a theoretical model, partially shields the atomic cloud from losses arising in the 461\,nm cooling cycle. We show a factor of two enhancement in the atom number for the bosonic isotopes $^{88}$Sr and $^{84}$Sr, and the fermionic isotope $^{87}$Sr, in good agreement with our model. Our scheme can be applied in the majority of strontium experiments without increasing the experimental complexity of the apparatus, since the employed 689\,nm transition is commonly used for further cooling. Our method should thus be beneficial to a broad range of quantum science and technology applications exploiting cold strontium atoms, and could be extended to other atomic species.
\end{abstract}
\maketitle

\section{\label{sec:introduction}Introduction}

Laser cooling of atomic gases has become a fundamental step for a large range of applications in quantum science and technology with cold and ultracold atoms~\cite{MetcalfStraten, schreck2021laser}. Starting from the alkali metals, the family of laser-cooled atomic elements has extended to more complex atomic species, as exemplified by alkaline-earth species such as strontium~\cite{katori1999magneto,xu2003cooling,stellmer2013laser}. The rich energy-level diagram of strontium, emerging from its two-valence electron structure, offers many exciting opportunities for quantum technologies. A well-known example is the extremely narrow clock transition of $^{87}\mathrm{Sr}$, which has become a reference in optical atomic clocks~\cite{Grotti2018, takamoto2020test, bothwell2022resolving}.
Other prominent applications include atomic gravimeters~\cite{ferrari2006long}, interferometers~\cite{hu2017atom,akatsuka2017optically}, superradiant lasers~\cite{norcia2016superradiance}, quantum simulation~\cite{martin2013, Schaefer2020}, and quantum computing~\cite{Madjarov2020, schine2022, barnes2022assembly}.
Strontium has also recently enabled the production of continuous Bose-Einstein condensates~\cite{chen2022continuous}, as well as studies of Rydberg~\cite{Camargo2018Rydberg, Bounds2018Rydberg} and molecular physics~\cite{McDonald2016}.
Most of these applications would benefit from increased atom number, in particular those in which strontium is brought to quantum degeneracy via evaporative cooling~\cite{stellmer2009boseeinstein,deescobar2009boseeinstein, DeSalvo2010}.
However, increasing the atom number typically requires optimized atomic sources~\cite{schioppo2012, ovchinnikov2008,yang2015high,barbiero2020sidebandenhanced, Chen2019strontiumbeam} or strategies to suppress the losses of the magneto-optical trap (MOT). These can greatly increase the complexity of the experimental apparatus and are at odds with the compactness required by quantum-technology applications.

Strontium MOTs suffer from high loss rates due to a cooling transition that is not fully closed and leads to the population of metastable states. Although these losses can be reduced by introducing repumping lasers~\cite{kurosu1992laser,dinneen1999cold,PoliPRA2005,mickelson2009repumping,stellmer2014reservoir,barker2015enhanced,ding2016narrow,moriya2018comparison,hu2019analyzing,zhang2020novel}, the limited efficiency of these schemes challenges a further increase of the atom number. One possible strategy to circumvent this problem is to prepare the cooled atoms in a reservoir state unaffected by such losses. For example, in the absence of repumping lasers strontium atoms slowly accumulate in a magnetically trapped metastable state during the MOT~\cite{katori2001laser,stellmer2009boseeinstein}. However, this approach comes at the expense of increasing the experimental cycle time. Schemes taking place on laser cooling timescales, analogous to dark-spot MOTs of alkali atoms~\cite{ketterle1993high,anderson1994reduction}, would be more favorable, but are complex to devise in the case of strontium.

In this work, we present an approach to increase the atom number of strontium MOTs on laser-cooling timescales that relies on populating a short-lived reservoir state. The key idea is to reduce the atomic losses arising during the MOT operating on the blue \SI{461}{\nm} transition of strontium by resonantly driving an additional transition that is fully cycling. The excited state of this transition acts as a reservoir state in which cold atoms are shielded from the MOT losses. In our case, this shielding transition is the red intercombination line at \SI{689}{\nm}, which is required for narrow-line cooling and is part of most strontium setups~\cite{katori1999magneto,xu2003cooling}. Hence, implementing our scheme does not add complexity to the experimental apparatus and is compatible with its miniaturization. We demonstrate an atom-number enhancement of a factor of two using this shielding method and show its versatility by applying it to both bosonic and fermionic strontium isotopes.

To present a comprehensive description and characterization of our shielding scheme, this paper is organized as follows:
Section~\ref{sec:laser cooling of strontium} introduces the relevant strontium energy levels and demonstrates the atom-number enhancement of the blue MOT.
In section~\ref{sec:model} we present a theoretical model of the observed atom-number enhancement and validate it by measuring the population dynamics between the involved atomic states.
Section~\ref{sec:frequency dependence} investigates the dependence of the enhancement on the frequency of the \SI{689}{\nm} beam.
Section~\ref{sec:robustness} shows the application of our scheme to different isotopes and repumping configurations.
Finally,
section~\ref{sec:conclusion and outlook} summarizes our findings and discusses its potential relevance for quantum science and technology applications.

\section{\label{sec:laser cooling of strontium}Enhancing the blue MOT with resonant red light}

\begin{figure}[!b]
\centering
\includegraphics[]{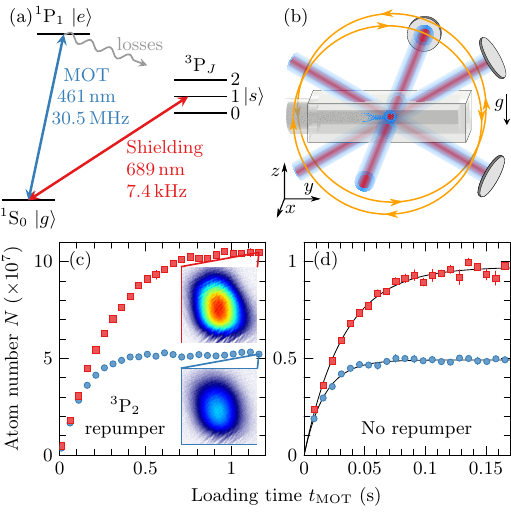}
\caption{(a) Energy level diagram of strontium, indicating the transitions relevant to this work (colored arrows).
The curved gray arrow indicates loss mechanisms from the \textsuperscript{1}P\textsubscript{1} state.
(b) Experimental setup. The shielding beams are overlapped with the blue MOT beams and both are retro-reflected. Coils in anti-Helmholtz configuration (yellow) provide the quadrupole magnetic field for the MOT.
(c)-(d) Loading curve of the $^{88}$Sr blue MOT with (red squares) and without (blue circles) the red resonant light. In (c) the \textsuperscript{3}P\textsubscript{2} state is continuously repumped, while this is not the case for (d). Insets in (c) are absorption images corresponding to the cloud with (top) and without (bottom) the red resonant light and a shared density color scale. Solid lines in (d) are fits using Eq.~\eqref{eq:rate equation} (see main text). Error bars denote one standard error of the mean (s.e.m.).}
\label{fig:energy level diagram}
\end{figure}

The optical transitions of atomic strontium relevant for laser cooling stem from its two valence electrons, and are displayed in Fig.~\ref{fig:energy level diagram}(a).
The blue transition, with $\lambda_{\mathrm{blue}}=\SI{461}{\nm}$, connects the \textsuperscript{1}S\textsubscript{0} ground state with the \textsuperscript{1}P\textsubscript{1} state. It has a broad linewidth of $\Gamma_{\textnormal{blue}}/2\pi=\SI{30.5}{\MHz}$ and is used to capture and cool the atoms to temperatures of few \si{\milli\K} in a blue MOT.
The red transition is the intercombination line \textsuperscript{1}S\textsubscript{0} $\rightarrow$ \textsuperscript{3}P\textsubscript{1} at $\lambda_{\mathrm{red}}=\SI{689}{\nm}$, with a linewidth of $\Gamma_{\textnormal{red}}/2\pi=\SI{7.4}{\kHz}$. This value is more than three orders of magnitude lower than $\Gamma_{\textnormal{blue}}$, and enables narrow-line laser cooling to temperatures of few \si{\ensuremath{\mu}\K} in a red MOT. Figure~\ref{fig:energy level diagram}(b) shows a sketch of our experimental setup, including the glass cell where the MOT is loaded, the blue and red beams used for laser cooling, and a pair of coils generating the MOT magnetic field. During the blue MOT, an additional laser beam at \SI{481}{nm} repumps the metastable \textsuperscript{3}P\textsubscript{2} state~\cite{ding2016narrow,hu2019analyzing}, which is populated through a leak channel of the blue transition. Further details on our experimental approach can be found in appendix~\ref{app:machine and losses}.

Our main result is the observation that driving the red transition on resonance during the blue MOT leads to a factor of two enhancement in the steady-state number of trapped atoms. This is shown in Fig.~\ref{fig:energy level diagram}(c), which displays the $^{88}$Sr MOT loading curves obtained via absorption imaging in the presence (red squares) and absence (blue circles) of red light.
While these measurements were taken by shining red resonant light in a MOT configuration (see appendix~\ref{app:machine and losses}), we have observed that the enhancement survives even in the presence of a single running-wave red beam.
Therefore, we conclude that the atom-number increase is not caused by a laser cooling effect on the red transition. Importantly, the narrower linewidth of the red transition results in a negligible radiation pressure with respect to that of the blue transition.

We attribute the observed enhancement to a modification of the loss mechanisms occurring in the blue MOT. Such losses can be of one-body ($\propto \gamma n$) or two-body ($\propto \beta n^2$, light-assisted) origin, where $\gamma$ and $\beta$ are the corresponding one-body and two-body loss coefficients, and $n$ is the atomic density. Due to imperfections in the repumping schemes, the two types of loss rates are known to be of similar magnitude in strontium MOTs~\cite{mickelson2009repumping}. This complicates the theoretical description of the MOT loading curves and the extraction of the loading and loss rates. To get a better insight into the origin of the enhancement, we simplify the system by analyzing the MOT in a regime dominated by one-body losses.
To this end, we switch off the laser that repumps the \textsuperscript{3}P\textsubscript{2} state, boosting the value of $\gamma$ and significantly reducing the atomic density.

Figure \ref{fig:energy level diagram}(d) shows that the enhancement effect persists in the limit of dominating one-body losses.
In this case, the loading process is described by a simple rate equation
\begin{equation}
    \frac{d N(t)}{d t} = L - \gamma N(t),
    \label{eq:rate equation}
\end{equation}
where $N(t)$ is the number of atoms in the MOT at time $t$ and the steady-state atom number is given by $N_{\textnormal{fin}}=L/\gamma$.
By fitting the experimental data of Fig. \ref{fig:energy level diagram}(d) with the solution of Eq. \eqref{eq:rate equation} we extract the loading and loss rates for each scenario.  In the absence of red light we obtain a bare loading rate of $L=\SI{2.9(1)e8}{\per\s}$, which is very similar to the value obtained with the red light $\tilde{L}=\SI{3.0(1)e8}{\per\s}$.
In contrast, the determined one-body loss rate is strongly affected and decreases from $\gamma=\SI{59(2)}{\per\s}$ to $\tilde{\gamma}=\SI{31(1)}{\per\s}$ in the presence of the red light.
These values, when compared with an estimated two-body loss rate, justify a posteriori our approximation of neglecting two-body losses. Indeed, based on the measured atomic density $n\approx\SI{2e10}{\per\cubic\cm}$ and the two-body loss coefficient reported in literature $\beta\approx \SI{2e-10}{\cubic\cm\per\s}$~\cite{dinneen1999cold,caires2004intensity}, we estimate a two-body loss rate $\beta n\approx \SI{4}{\per\second}\ll\gamma,\tilde{\gamma}$.

As a result of the reduced loss rate in the presence of red light $\tilde{\gamma}$, the steady-state atom number increases from $N_{\textnormal{fin}}=\num{4.92(3)e6}$ to $\tilde{N}_{\textnormal{fin}}=\num{9.7(1)e6}$, resulting in an enhancement of the MOT atom number by a factor $\eta=\num{1.97(1)}$.
This enhancement takes place in both repumped and non-repumped MOTs, for which the losses have very different origins.
Moreover, removing the repumper significantly reduces the density, from $\approx \SI{1.3e11}{\per\cubic\cm}$ to $\approx \SI{2e10}{\per\cubic\cm}$, but the enhancement remains unchanged.
From these results, we conclude that it is possible to shield atoms from losses across a wide range of densities.

\section{\label{sec:model}Theoretical model and MOT fluorescence dynamics}

To understand the origin of the observed loss reduction, we provide a level-based interpretation of the MOT-loading rate Eq.~\eqref{eq:rate equation} by considering the internal structure of strontium, restricting our analysis to states \ensuremath{\left\vert g \right\rangle}, \ensuremath{\left\vert e \right\rangle} and \ensuremath{\left\vert s \right\rangle} (respectively \textsuperscript{1}S\textsubscript{0}, \textsuperscript{1}P\textsubscript{1}, \textsuperscript{3}P\textsubscript{1}, see Fig.~\ref{fig:energy level diagram}(a)).
We assume that atoms in \ensuremath{\left\vert g \right\rangle} experience losses $\gamma$ from their continuous excitation to \ensuremath{\left\vert e \right\rangle}, while losses associated to the \ensuremath{\left\vert g \right\rangle}~$\to$~\ensuremath{\left\vert s \right\rangle} transition are negligible.
The lack of losses in the \ensuremath{\left\vert s \right\rangle} state is a crucial requirement for our model to apply.
In a conventional MOT, most of the atomic population is in the ground state $N_g \gg N_e$, such that the total atom number can be approximated by $N\approx N_g$.
However, when resonantly driving the \ensuremath{\left\vert g \right\rangle}~$\to$~\ensuremath{\left\vert s \right\rangle} transition, a fraction $\alpha$ of the atoms remains excited in state \ensuremath{\left\vert s \right\rangle} and the total atom number becomes $N\approx N_g+N_s$, with $N_s=\alpha N$ and $N_g=(1-\alpha) N$.
Since only atoms in \ensuremath{\left\vert g \right\rangle} experience losses via their excitation to \ensuremath{\left\vert e \right\rangle}, we can rewrite Eq. \eqref{eq:rate equation} as
\begin{equation}
    \frac{d N(t)}{d t} = L - (1-\alpha) \gamma N(t).
    \label{eq:rate equation 2}
\end{equation}
Therefore, the effective losses in the presence of the shielding beam are reduced to $\tilde{\gamma}=(1-\alpha)\gamma$. This leads to an increase of the steady-state total atom number, $\tilde{N}_{\textnormal{fin}}=L/(1-\alpha)\gamma$. While for simplicity the description above considered only one-body losses, in appendix \ref{app:light-assisted collissions} we show that when including two-body losses similar arguments apply.

We experimentally confirm our model and its expectations by monitoring the population of the \ensuremath{\left\vert g \right\rangle} and \ensuremath{\left\vert e \right\rangle} states during the MOT dynamics. To this end, we compare measurements performed in two configurations. On the one hand, we extract the total atom number from absorption imaging after switching off all beams, as was done in Fig.~\ref{fig:energy level diagram}. On the other hand we continuously track the fluorescence emitted by the MOT on the blue \ensuremath{\left\vert e \right\rangle}~$\to$~\ensuremath{\left\vert g \right\rangle} transition, which we collect using a narrow bandpass filter at 461\,nm and a photomultiplier. This fluorescence signal is proportional to $N_g$, since atoms stored in \ensuremath{\left\vert s \right\rangle} do not scatter blue photons. Thus, comparing absorption and fluorescence measurements provides insight into the population of the involved atomic states.

\begin{figure}[!t]
\centering
\includegraphics[]{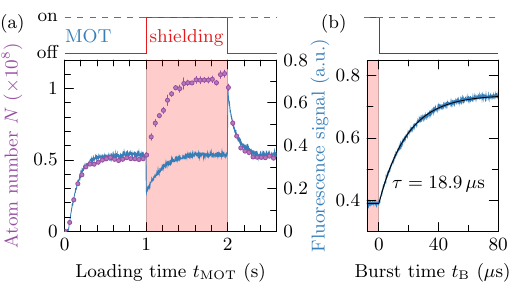}
\caption{(a) Measured $^{88}$Sr atom number (violet) and $461$ nm fluorescence signal (proportional only to $N_g$, blue) for different MOT loading times $t_{\textnormal{MOT}}$. Top: Scheme of the experimental sequence. Switching on the red light for $t_\mathrm{MOT}=\SI{1}{\second}$ (red-shaded region) triggers fluorescence and atom number dynamics due to the shielding from losses (see text). Error bars denote one s.e.m. (b) Increase of fluorescence after the red light switch-off at $t_\mathrm{MOT}=\SI{2}{\second}$, note the very different timescale. Black solid line: exponential fit to the data with an extracted time constant $\tau=\SI{18.9(1)}{\ensuremath{\mu}s}$, compatible with the lifetime of the \ensuremath{\left\vert s \right\rangle} state $\tau_s=\SI{21.28(3)}{\ensuremath{\mu}s}$~\cite{nicholson2015}.}
\label{fig:fluorescence}
\end{figure}%

Figure \ref{fig:fluorescence}(a) shows the total atom number and the fluorescence signal recorded during the loading of the blue MOT. These measurements are taken in the presence of the \textsuperscript{3}P\textsubscript{2} repumper, i.e., the same configuration as in Fig.~\ref{fig:energy level diagram}(c).
During the initial stage, without the red shielding beam, both atom number and fluorescence grow monotonically as a function of the loading time $t_\mathrm{MOT}$, and saturate to a steady-state value. At $t_\mathrm{MOT} = \SI{1}{s}$, we abruptly switch on the red light on resonance with the \ensuremath{\left\vert g \right\rangle}~$\to$~\ensuremath{\left\vert s \right\rangle} transition. We observe that this leads to an immediate drop of the fluorescence signal to approximately half of its steady-state value, while the measured total atom number remains the same. Our interpretation is that the red light saturates the red transition, equalizing the populations in the \ensuremath{\left\vert g \right\rangle} and \ensuremath{\left\vert s \right\rangle} states on a timescale which is much shorter than the MOT loading timescale. Subsequently, approximately half of the atoms are in \ensuremath{\left\vert s \right\rangle}. Since in this state they do not experience the losses of the blue cooling cycle, the atom number and fluorescence signal of the MOT increases again. We indeed observe a second loading stage until a new steady-state is reached.
There, the fluorescence signal is comparable to the one obtained without red light, but the total atom number is approximately doubled. The observed atom-number enhancement $\eta=2.08(15)$ agrees with the fraction of atoms occupying the \ensuremath{\left\vert s \right\rangle} state inferred from the fluorescence signal $\alpha=0.47(1)$. Moreover, from the loss rates fitted in Fig.~\ref{fig:energy level diagram}(d), we obtain $\alpha=1-\tilde{\gamma}/\gamma= 0.47(2)$. These values are consistent with a population of state \ensuremath{\left\vert s \right\rangle} corresponding to half of the total atom number.

To further confirm our interpretation, once the steady state is reached at $t_\mathrm{MOT} = \SI{2}{s}$, we switch the red light back off. As shown in Fig.~\ref{fig:fluorescence}(b), we observe a rapid rise in the fluorescence signal to approximately twice its steady-state value, which we attribute to the fast decay of the $\ensuremath{\left\vert s \right\rangle}$ population to the ground state $\ensuremath{\left\vert g \right\rangle}$. An exponential fit to the fluorescence signal yields a time constant $\tau=\SI{18.9(1)}{\mu s}$, which is comparable to the lifetime of the \ensuremath{\left\vert s \right\rangle} state $\tau_s=\SI{21.28(3)}{\mu s}$~\cite{nicholson2015}. At long times after this fluorescence spike, the MOT losses lead to a much slower decay in both fluorescence and atom number. As shown in Fig.~\ref{fig:fluorescence}(a), they eventually bring the MOT to its steady-state value in the absence of red light.
In conclusion, the population dynamics reported in this section provides further evidence that the atom-number enhancement is caused by a fraction $\alpha\approx0.5$ of the atoms populating \ensuremath{\left\vert s \right\rangle}, which is shielded from the losses associated to the blue cooling light.

\section{\label{sec:frequency dependence}Enhancement spectrum}

\begin{figure}[!t]
\centering
\includegraphics[]{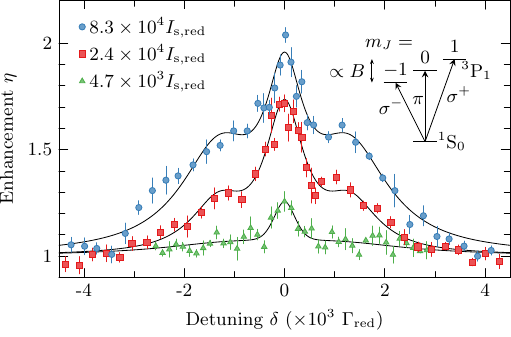}
\caption{Enhancement factor $\eta$ for $^{88}$Sr as a function of the detuning from resonance $\delta$, measured for different intensities of the red shielding beam. Black lines: empirical fits to the experimental profiles (see main text). Inset: energy-level diagram of the \textsuperscript{1}S\textsubscript{0}~$\to$~\textsuperscript{3}P\textsubscript{1} transition displaying the Zeeman splitting of the different $m_J$ Zeeman substates, which are responsible for the characteristic shape of the measured spectra. Error bars denote one s.e.m.
}
\label{fig:enhancement}
\end{figure}

To gain further insight, we investigate the frequency dependence of our red shielding scheme. To this end, we determine the enhancement factor $\eta$ at different frequencies by consecutively preparing an atomic cloud with and without the red shielding beams and dividing their respective atom numbers.
The measurement is performed repumping the \textsuperscript{3}P\textsubscript{2}  state, in the same configuration as Fig.~\ref{fig:energy level diagram}(c).
Figure~\ref{fig:enhancement} shows the enhancement spectra measured as a function of the detuning $\delta$ from the red resonance, for different intensities of the red shielding beam. We express the latter in terms of the saturation parameter $s_{\textnormal{red}}=I/I_{\textnormal{s,red}}$, where $I_{\textnormal{s,red}}=\SI{2.96}{\mu W/cm^2}$ is the saturation intensity of the shielding transition. Experimentally, we observe that all spectra display a characteristic shape, consisting of a narrow peak at $\delta=0$ surrounded by a broad background.

We attribute this shape to the addressing of the different Zeeman substates $m_J$ of \textsuperscript{3}P\textsubscript{1}, which become resonant at different values of the detuning $\delta$.
On one hand, the $\pi$ transition to state $m_J=0$, which is magnetically insensitive, is only affected by Doppler and intensity broadening.
On the other hand, the $\sigma^\pm$ transitions to states $m_J=\pm1$ are magnetically sensitive, and also experience Zeeman broadening due to the MOT quadrupole magnetic field. Thus, we assign the narrow peak of the spectra to the $\pi$ transition. It can be addressed because, although the shielding beams are circularly polarized,  their polarization with respect to the transition is not well defined at the zero of the quadrupole magnetic field. We identify the broad background with two partially overlapping side peaks that correspond to the $\sigma^\pm$ transitions.

To confirm this interpretation, we fit the experimental data points with an empirical combination of 3 Voigt profiles, a central one and two identical side ones.
For each Voigt profile, one contribution to the width comes from Doppler broadening. It is determined from an independent temperature measurement and given by
 $\sqrt{k_{\textnormal{B}} T /m}/\lambda_{\textnormal{red}}\approx\SI{1.38}{\MHz}$, where $T\approx\SI{9.6}{\milli\K}$ is the temperature, $k_{\textnormal{B}}$ is the Boltzmann constant and $m$ the mass of $^{88}$Sr.
The second contribution to the width, which comes from the intensity and Zeeman broadening, is left as a free parameter.
For the highest saturation parameter $s_{\textnormal{red}}\approx\num{8.3e4}$, the fit yields $\Delta f_{\pi}=\SI{2.4(1)}{\MHz}$ for the central peak and $\Delta f_{\sigma}=\SI{6.6(3)}{\MHz}$ for the side peaks. The measured value of $\Delta f_\pi$ is in good agreement with the linewidth of the transition when taking into account the intensity broadening, $\sqrt{1+s_{\textnormal{red}}}\,\Gamma_{\textnormal{red}}/2\pi \approx \SI{2.13}{\MHz}$. The measured value of $\Delta f_\sigma$ is comparable with the expected Zeeman broadening of the transition $1.5\mu_{\textnormal{B}} B' r_{\textnormal{MOT}} / h \approx \SI{7.1}{\MHz}$, where $\mu_B$ is the Bohr magneton, and we have considered the magnetic field gradient $B'\approx \SI{42}{G\per\cm}$ and the measured size of the MOT $r_{\textnormal{MOT}}\approx \SI{0.8}{\mm}$ along the weak magnetic field directions. For the other saturation parameters, the fits yield comparable results.

In conclusion, our measurements of the enhancement spectrum are in good agreement with our model and show that the shielding is a resonant effect. Since the width of the intensity-broadened profile is larger than the Doppler broadening of the cloud, we are able to saturate the $\pi$ transition in a large fraction of the MOT. This explains the factor of two observed in the enhancement. Saturating the red transition is indeed required to reach optimal shielding. Note that, for simplicity, the model introduced in Sec.~\ref{sec:model} neglects the Zeeman substructure. This assumption is justified by the fact that the shielding transition is driven on resonance, mostly addressing the $\pi$ transition.

\section{\label{sec:robustness}Shielding for different isotopes and repumping configurations}

To show the generality of the shielding scheme, in this section we apply it to a different repumping configuration and to the other isotopes of strontium. In most strontium experiments, the MOT repumping scheme exploits two lasers. One repumps the \textsuperscript{3}P\textsubscript{2} state, as discussed in the previous sections, and a second one repumps the \textsuperscript{3}P\textsubscript{0} (clock) state, which gets populated if the \textsuperscript{3}P\textsubscript{2} repumping cycle is not closed \cite{dinneen1999cold}. Given that the 481\,nm transition that we use to repump \textsuperscript{3}P\textsubscript{2} is only nearly closed~\cite{hu2019analyzing}, we observe that introducing an additional repumper of the \textsuperscript{3}P\textsubscript{0} state at 679\,nm (see details in appendix~\ref{app:machine and losses}) removes a residual leakage channel and increases the MOT atom number.

\begin{figure}[!h]
\centering
\includegraphics[]{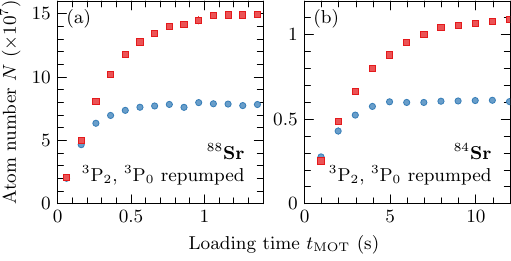}
\caption{Loading curves of the blue MOT for two bosonic isotopes, measured when repumping the metastable \textsuperscript{3}P\textsubscript{2} as well as the clock \textsuperscript{3}P\textsubscript{0} state. The data shows loading with (red squares) and without (blue circles) the shielding beams. The shielding effect due to the resonant driving of the red transition is comparable for $^{88}\mathrm{Sr}$ (a) and $^{84}\mathrm{Sr}$ (b), albeit with a lower absolute atom number due to the difference in natural abundances. Error bars denote one s.e.m.}
\label{fig:loading curves repumpers}
\end{figure}%
In Fig.~\ref{fig:loading curves repumpers}(a), we investigate the loading of the $^{88}\mathrm{Sr}$ isotope in the presence of this additional repumper, and show that the shielding effect persists and yields a comparable enhancement factor $\eta=1.94(1)$.
This brings the highest atom number we achieve for this isotope, of $N \approx 1.5\times 10^8$. We obtain a similar enhancement factor for the second most abundant bosonic isotope $^{86}\mathrm{Sr}$. We also probe this effect for the least abundant bosonic isotope $^{84}\mathrm{Sr}$, which is preferred for Bose-Einstein condensation due to its favorable scattering properties \cite{stein2008, deEscobar2008}. Here, the resonant driving of the intercombination line increases the atom number by a factor of $\eta = 1.85(2)$, as shown in Fig.~\ref{fig:loading curves repumpers}(b). Importantly, this enhancement in the atom number is kept through the following steps of our experiment, up to the loading into an optical dipole trap. The shielding scheme therefore improves the efficiency of the evaporative cooling process and greatly facilitates the entrance into the quantum-degenerate regime, allowing us to produce pure Bose-Einstein condensates with more than $3\times 10^{5}$ atoms.

\begin{figure}[!t]
\centering
\includegraphics[]{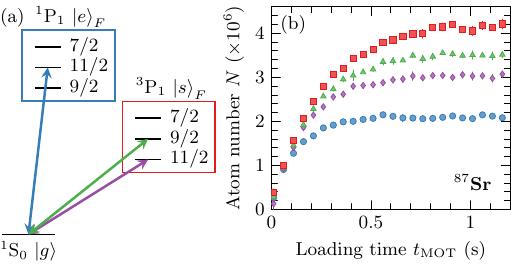}
\caption{(a) Hyperfine structure of the fermionic isotope $^{87}$Sr relevant for this work. (b) Loading curves of fermions in the blue MOT, repumping only the \textsuperscript{3}P\textsubscript{2} state. The steady state atom number increases when shielding atoms in the \ensuremath{\left\vert s \right\rangle}$_{F=11/2}$ (purple diamonds) or in the \ensuremath{\left\vert s \right\rangle}$_{F=9/2}$ (green triangles) hyperfine states. Addressing the two transitions \ensuremath{\left\vert s \right\rangle}$_{F=11/2}$ and \ensuremath{\left\vert s \right\rangle}$_{F=9/2}$ simultaneously (red squares) enhances the total atom number compared to no shielding light at all (blue points) by a factor of $2.01(1)$. Error bars denote one s.e.m.}
\label{fig:loading curves fermions}
\end{figure}

Finally, we  also demonstrate the shielding effect for the fermionic isotope $^{87}\mathrm{Sr}$. This isotope is particularly relevant for optical atomic clocks. It has a nuclear spin of $I=9/2$, resulting in a hyperfine structure with quantum numbers $F$ and $m_F$ for both the ground and the excited states (see Fig.~\ref{fig:loading curves fermions}(a)).
This brings complexity to the shielding scheme, reducing its efficiency due to the lack of a magnetically insensitive $\pi$ transition. A positive aspect is, however, that all three hyperfine states of \textsuperscript{3}P\textsubscript{1} can be exploited for shielding.
In Fig.~\ref{fig:loading curves fermions}(b), we study the enhancement when addressing the $F=11/2$ and the $F=9/2$ states.
Lasers addressing the two transitions are readily available in fermionic strontium setups, because they correspond to the red MOT and stirring beam \cite{mukaiyama2003recoillimited}. Our measurements show that the $F=11/2$ shielding enhances the atom number by a factor of $1.45(1)$, while addressing instead the $F=9/2$ state gives an enhancement of $\eta=1.69(1)$.
The larger enhancement while driving the $F=9/2$ transition is explained by its small Zeeman broadening, which stems from a factor of $4.5$  smaller Land\'e $g$-factor than the one of $F=11/2$. Thus, the laser addresses a larger fraction of the cloud.
When driving the two transitions simultaneously, we combine the individual enhancements to a total $\eta =  2.01(1)$.

Table~\ref{tab:enhancements} shows the similarity of the enhancement factors reported in sections \ref{sec:laser cooling of strontium} and \ref{sec:robustness}. The consistency of all measured enhancement values indicates the generality of our shielding scheme, which applies to different repumping configurations and to isotopes with different quantum statistics and natural abundances.
In appendix \ref{app:diffatoms} we summarize general criteria for successfully applying the shielding and discuss its potential application to other atomic species.

\begin{table}[!htb]
\centering
\caption{Atom-number enhancements $\eta$ measured in different repumping configurations, and for bosonic ($^{88}$Sr, $^{84}$Sr) and fermionic ($^{87}$Sr) isotopes of different  natural abundances \cite{moore1982absolute}. The enhancement factors are extracted from exponential fits of the loading curves. Uncertainties come from the standard deviation errors of the fit parameters. For the fermionic isotope, we address two shielding transitions (see main text).}
\label{tab:enhancements}
\begin{ruledtabular}
\begin{tabular}{@{}lcccc@{}}

                                        & Abundance ($\%$) & Repumping &  $\eta$  & Data \\
                                        \midrule
$^{88}$Sr            & \num{82.58} & None & \num{1.97(1)} & Fig.~\ref{fig:energy level diagram}(d)\\
                     &  & \textsuperscript{3}P\textsubscript{2} & \num{2.08(1)} &Fig.~\ref{fig:energy level diagram}(c)\\
                     &  & \textsuperscript{3}P\textsubscript{2}, \textsuperscript{3}P\textsubscript{0} & \num{1.94(1)} &Fig.~\ref{fig:loading curves repumpers}(a)\\
\midrule
$^{84}$Sr            & \num{0.56}  & \textsuperscript{3}P\textsubscript{2}, \textsuperscript{3}P\textsubscript{0} & \num{1.85(2)}  & Fig.~\ref{fig:loading curves repumpers}(b)\\
\midrule
$^{87}$Sr            & \num{7.00}  & \textsuperscript{3}P\textsubscript{2} & \num{2.01(1)} & Fig.~\ref{fig:loading curves fermions}(b)\\
\end{tabular}
\end{ruledtabular}
\end{table}

\section{\label{sec:conclusion and outlook}Conclusion and outlook}

In this work we have presented a scheme to increase the steady-state atom number in strontium blue MOTs by resonantly driving the red intercombination line.
Partially populating the \textsuperscript{3}P\textsubscript{1} excited state effectively shields the MOT from losses caused by the blue cooling light, making it act as a short-lived reservoir state. Saturating the transition leads to a factor of two  enhancement in the steady-state atom number.
We have developed a simple theoretical model which captures the main features of the shielding effect. It provides insight into the state population dynamics and the shape of the enhancement spectrum that we have observed experimentally.
Moreover, we have seen that the enhancement takes place in MOTs with vastly different atomic densities, indicating that the shielding effect persists regardless of the loss regime.
While in our experiments we focus on a particular choice of repumping lasers, the enhancement will apply similarly to experiments with other repumping schemes.
Moreover, we have demonstrated the generality of the shielding method by applying it to the bosonic $^{88}$Sr and $^{84}$Sr, as well as the fermionic  $^{87}$Sr.
We have also shown that simultaneously addressing two reservoir states leads to a combined gain. Finally, we have observed that the atom-number enhancement persists through the subsequent cooling stages of our experiment, including a narrow-line red MOT and the loading of the atoms into an optical dipole trap, and greatly facilitates achieving quantum degeneracy.

Thanks to its simplicity and generality, we expect our shielding scheme to benefit a broad spectrum of quantum science and technology experiments involving laser-cooled strontium atoms. This is especially the case for those applications that are critical in the atom number, such as superradiant lasers~\cite{norcia2016superradiance}, quantum simulators~\cite{Schaefer2020} and atom interferometers~\cite{hu2017atom,akatsuka2017optically}, including some recent large-scale projects~\cite{badurina2020aion,el-neaj2020aedge}. Optical atomic clocks could also profit in certain next-generation approaches, such as quantum-degenerate optical lattice clocks~\cite{campbell2017a}, continuous Ramsey clocks~\cite{katori2021longitudinal} or interleaved cold-atom clocks~\cite{takamoto2011frequency}.
Existing experiments can readily benefit from our scheme, since it requires no additional changes in the setups and it is compatible with the miniaturization of experiments~\cite{Grotti2018, takamoto2020test,kwon2023jetloaded}.
Going beyond our work, we expect that our shielding method will find application in experiments with other atomic species.

\begin{acknowledgments}
We thank D. Jacobs for his contributions to the design and construction of the experimental apparatus, C. S. Chisholm, S. Hirthe and R. Ramos for a careful reading of the manuscript, and the ICFO Quantum Gases Experimental group for discussions. We are grateful to J. C. Cifuentes, C. Dengra, X. Menino and the corresponding ICFO facilities for technical support. We acknowledge funding from the European Union (PASQuanS2.1 project No. 101113690 and DAALI project No. 899275), MCIN/AEI/10.13039/501100011033 (LIGAS project PID2020-112687GB-C21, DYNAMITE QuantERA project PCI2022-132919 with funding from European Union NextGenerationEU, Equipamiento Científico Técnico EQC2018-005001-P and EQC2019-005699-P, Severo Ochoa CEX2019-000910-S, and PRTR-C17.I1 with funding from European Union NextGenerationEU and Generalitat de Catalunya), Fundación Ramón Areces (project CODEC), Fundació Cellex, Fundació Mir-Puig, and Generalitat de Catalunya (ERDF Operational Program of Catalunya, Project QUASI-CAT/QuantumCat Ref. No. 001-P-001644, Departament de Recerca i Universitats 2021-SGR-01448, AGAUR and CERCA program).
J.H. acknowledges support from the European Union (Marie Sk\l{}odowska-Curie–713729), S.B. from MCIN/AEI/10.13039/501100011033 and ESF (PRE2020-094414), A.R. from the MCIN/AEI/10.13039/501100011033 (Juan de la Cierva Formación FJC2020-043086-I), and V.M. from the Beatriu de Pinós Program and the Ministry of Research and Universities of the Government of Catalonia (2019-BP-00228).
\end{acknowledgments}

\appendix

\section{Experimental setup and repumping scheme}
\label{app:machine and losses}

Our experimental apparatus consists of a glass cell with ultrahigh vacuum in which we load the blue MOT from an atomic beam. This beam originates from a transversely loaded two-dimensional MOT fed by an oven and a short Zeeman slower, inspired by the compact design of Refs.~\cite{lamporesi2013compact,nosske2017twodimensional, barbiero2020sidebandenhanced}. As shown in Fig.~\ref{fig:energy level diagram}(b), the blue MOT uses three retro-reflected $\sigma^+/\sigma^-$ polarized laser beams at $\lambda_{\textnormal{blue}}$,
and a magnetic quadrupole field generated by two Bitter coils~\cite{sabulsky2013efficient} in anti-Helmholtz configuration, with a magnetic field gradient of $B'=\SI{42}{G\per\cm}$ along the weak-field directions.
The blue MOT beams are overlapped with three retro-reflected $\sigma^+/\sigma^-$ polarized red beams at $\lambda_{\textnormal{red}}$ for subsequent narrow-line laser cooling in a red MOT and to enhance the atom number in the blue MOT.
To ensure a laser linewidth well below $\Gamma_{\mathrm{red}}$, the red laser is stabilized to an ultra-low expansion glass cavity with finesse $\mathcal{F}\approx \num{1.9e5}$.

\begin{figure}[!t]
\centering
\includegraphics[]{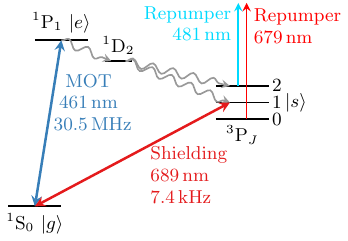}
\caption{Energy level diagram of strontium, including the cooling, shielding and repumping transitions relevant to this work (colored arrows).
Curved gray arrows indicate optical leakage channels. The \SI{481}{nm} laser repumps the population of the \textsuperscript{3}P\textsubscript{2} state, but leads to a finite population in the \textsuperscript{3}P\textsubscript{0} state, which can be repumped by the additional \SI{679}{nm} laser. }
\label{fig:energy level diagram appendix}
\end{figure}

In addition to the blue and red lasers, we can also use two repumper lasers in our MOT, which address single-body loss mechanisms associated to the blue transition, as depicted in Fig. \ref{fig:energy level diagram appendix}.
In particular, during the blue MOT atoms in \textsuperscript{1}P\textsubscript{1} decay with a probability of $1:20000$~\cite{cooper2018alkalineearth} into the \textsuperscript{1}D\textsubscript{2} state. This state has a lifetime $\tau_{\textnormal{D}}=\SI{330}{\mu s}$ and decay channels to the \textsuperscript{3}P\textsubscript{1} and  \textsuperscript{3}P\textsubscript{2} states. While atoms in \textsuperscript{3}P\textsubscript{1} decay back to the ground state, \textsuperscript{3}P\textsubscript{2} is a metastable state with a lifetime longer than experimental timescales. Therefore, atoms decaying to it are lost from the cooling cycle, which limits the atom number in the blue MOT.
In our experiment we continuosly repump the atoms falling to the metastable \textsuperscript{3}P\textsubscript{2} state via the \SI{481}{\nm} transition to the doubly excited \textsuperscript{3}P\textsubscript{2} state~\cite{ding2016narrow,hu2019analyzing}.
The limiting factor of this method are the remaining loss mechanisms in the MOT. They correspond to imperfections in the repumping process, which can lead to the population of another metastable state \textsuperscript{3}P\textsubscript{0}~\cite{stellmer2014reservoir,hu2019analyzing}, or to the escape of the atoms from the MOT region while they populate the untrapped substates of \textsuperscript{1}D\textsubscript{2}~\cite{moriya2018comparison}.
In section~\ref{sec:robustness} we introduce a second repumping laser at \SI{679}{\nm} which repumps the \textsuperscript{3}P\textsubscript{0} state~\cite{dinneen1999cold}, closing the repumping cycle.

When operating the MOT using the \textsuperscript{3}P\textsubscript{2} repumper and the red shielding beam, with the laser beam intensities and detunings summarized in Table \ref{tab:beam parameters}, we obtain in the steady state an atomic cloud of $^{88}$Sr with $N\approx\num{1e8}$ atoms.
\begin{table}[!b]
\centering
\caption{Parameters for the blue MOT beams and the red shielding beams: single-beam power $P$, $1/e^2$ Gaussian beam waist $w$, on-resonance saturation parameter $s$ (for 6 beams in total), and detuning from the corresponding resonance $\Delta$ in units of its linewidth $\Gamma$. \label{tab:beam parameters}}
\begin{ruledtabular}
\begin{tabular}{lcccc}
            & \hspace{0.3cm}$P$ (mW)\hspace{0.3cm}   & \hspace{0.2cm}$w$ (mm)\hspace{0.2cm}          &\hspace{0.3cm} $s$\hspace{0.3cm}               & \hspace{0.2cm} $\Delta$ $(\Gamma)$                \\ \hline
Blue beams  & 3.5                & 3.88               & \num{2.2}                      & $-2.2$    \\
Red beams   & 3.0                & 2.16               & \num{8.3e4}                    & 0                       \\
\end{tabular}
\end{ruledtabular}
\end{table}
The temperature of the cloud, extracted from a time-of-flight measurement, is of $T\approx\SI{9.6}{\milli\K}$, and the cloud has a Gaussian rms width $r_{\textnormal{MOT}}\approx\SI{0.8}{\mm}$ along the weak-field directions. For the absorption imaging measurements in the main text we take a time-of-flight of $t_{\textnormal{TOF}}=\SI{1.1}{\ms}$ after switching off the MOT and red laser beams.

\section{Shielding in the presence of light-assisted collisions}
\label{app:light-assisted collissions}

In this appendix we adapt the model of section~\ref{sec:model} to include light-assisted losses.
This type of losses depends on the density of the MOT.
Following Ref.~\cite{weiner1999experiments}, we consider two limits of MOT densities, namely the low- and the high-density limits.

In the high-density limit, radiation trapping effects maintain a constant MOT density $n_c$. Such light-assisted losses become effectively single-body:
\begin{equation}
    \frac{dN(t)}{dt}=L-\left(\gamma +\beta n_c\right) N(t),
\end{equation}
where $\beta$ is the two-body loss coefficient.
Thus, the model in section~\ref{sec:model} can be directly applied by replacing $\gamma$ by $\gamma+\beta n_c$ and introducing an overall loss reduction factor $1-\alpha$, where $\alpha$ can be as large as $0.5$ for a single transition.
This gives an effective loss coefficient $\left(\gamma+\beta n_c\right)\left(1-\alpha\right)$.
For simplicity, we assume that the shielding does not cause a change in the density $n_c$.
In reality, the shielding light may lead to a decrease of $n_c$ due to radiation trapping of the shielding light photons.
This can lead to further decrease of the loss coefficient.
We stress that the decrease of losses due to the shielding ($1-\alpha$) and due to the decrease of density ($\beta n_c$) are different processes.
The combination of these two processes may potentially lead to an enhancement factor greater than 2.

In contrast, in a low-density MOT the volume remains constant while the density grows with the atom number.
Then, losses have two-body character and the MOT rate equation becomes
\begin{equation}
    \frac{dN(t)}{dt}=L-\gamma N(t)-\beta' N(t)^2,
\end{equation}
where $\beta'=\beta/\pi^{3/2}V$, and $V$ is the volume of the trap.
The steady-state atom number is
\begin{equation}
   N_{\textnormal{fin}}=\frac{\sqrt{\gamma^2+4\beta' L}-\gamma}{2\beta'}.
   \label{eq:Nss:tbl}
\end{equation}
The shielding appears as a reduction of the loss coefficients: $\gamma(1-\alpha)$ and $\beta'(1-\alpha)^2$.
Then Eq. \eqref{eq:Nss:tbl} may be rewritten as
\begin{equation}
     N_{\textnormal{fin}}=\frac{\sqrt{\gamma^2+4\beta' L}-\gamma}{2\beta'}\frac{1}{1-\alpha}.
\end{equation}
Thus, the enhancement appears as well in the case of two-body losses in a low-density MOT.
From our derivation, we conclude that the shielding scheme will enhance the steady-state atom number by effectively reducing the losses, regardless of whether they have a single-body or two-body nature. However, it will not reduce losses that affect all states equally, such as those stemming from background-gas collisions.

\section{\label{app:diffatoms} Shielding in other atomic species}

This appendix examines the possibility that other atomic species exhibit a shielding effect similar to the one observed in this work for strontium.
We identify three requirements for the shielding:
(i) the resonant coupling to a reservoir state does not introduce additional losses;
(ii) the radiation pressure of the shielding beam is smaller than that of the MOT beams, i.e., $\Gamma_{s}/\lambda_{s}\ll\Gamma_{e}/\lambda_{e}$. This condition can be relaxed in radiation-pressure balanced configurations, e.g. when using retro-reflected shielding beams;
(iii) the Doppler broadening of the shielding transition can be compensated by its intensity broadening. This precludes the use of clock or very long-lived states. Indeed, setting the Doppler broadening equal to the intensity broadening, we obtain a characteristic intensity which is in practice unreachable for ultranarrow transitions.
\begin{table}[!h]
\centering
\caption{Combinations of MOT/shielding transitions for alkaline-earth-like species }
\label{tab:species:ea}
\begin{ruledtabular}
\begin{tabular}{@{}lccccc@{}}

    & $\lambda_{e}$\,(nm) \hspace{0.0cm} & $\Gamma_{e}/2\pi$\,(MHz) \hspace{0.0cm} & $\lambda_{s}$\,(nm)  \hspace{0.0cm} & $\Gamma_{s}/2\pi$\,(kHz) \hspace{-0.1cm} & Ref.  \\ \hline
Sr        & \num{461} & \num{30} & \num{689}  & \num{7.4} & this work\\
Mg        & \num{285} & \num{79} &\num{457}   & \num{0.036} & \cite{riedmann2012beating}  \\
Ca  & \num{423} &\num{35}  &\num{658}   &\num{0.37} &\cite{kraft2009bose} \\
Cd  & \num{229} & \num{91} & \num{325}  & \num{70}  &\cite{brickman2007magneto} \\
Hg  & \num{185} & \num{30} &  \num{254} & \num{1270} &\cite{hachisu2008trapping} \\
Yb  & \num{399} & \num{28} &\num{556}   & \num{181} &\cite{gothe2019continuous}
\end{tabular}
\end{ruledtabular}
\end{table}

\begin{table}[!b]
\centering
\caption{Combinations of MOT/shielding transitions for rare-earth elements}
\label{tab:species:lanth}
\begin{ruledtabular}
\begin{tabular}{@{}lllll@{}}
~\hspace{1.2cm} & $\lambda_{e}$\,(nm)\hspace{0.2cm} & $\Gamma_{e}/2\pi$\,(MHz) \hspace{0.2cm}& $\lambda_{s}$\,(nm)  \hspace{0.2cm}& $\Gamma_{s}/2\pi$\,(kHz)  \\  \hline
Dy & \num{421}  & \num{32} & \num{598}  & \num{140}  \\
\cite{lu2011spectroscopy,petersen2020spectroscopy} & \num{598}  & \num{0.14} & \num{626}  & \num{135}   \\
   &  626 & 0.135 & 741  & 1.8   \\
   &   &  & 1001  & 1.8$\cdot10^{-3}$    \\ \hline
Ho & 411  & 33 & 599  & 147  \\
\cite{miao2014magneto,stefanska2020investigations} & 599  & 0.147 & 608  & 39   \\
   &   &  & 661  & 15   \\
   &   &  & 545  & 5    \\ \hline
Er & 401  & 30 & 583  & 186  \\
\cite{ban2005laser,patscheider2021observation} & 583  & 0.186 & 631  & 28  \\
 & & & 841  & 8  \\
 & & & \num{1299}  & \num{0.9}$\cdot10^{-3}$  \\ \midrule
 Tm & 410  & 10 & 531  & 530  \\
\cite{sukachev2010magneto} & 531  & 0.53 & 1140  & \num{1}$\cdot10^{-3}$  \\
\end{tabular}
\end{ruledtabular}
\end{table}

In the following, we provide combinations of MOT and shielding transitions for different atomic species which meet these requirements.
For alkaline-earth-like atoms we summarize such combinations in Table~\ref{tab:species:ea} and cite relevant experiments.
Rare-earth elements feature a number of cycling transitions convenient for both laser cooling and shielding.
In Table~\ref{tab:species:lanth} we list those that may be convenient for shielding, sorted by decreasing linewidth. According to requirement (ii) the MOT transition should be wider than the shielding transition.

Finally, in alkali atoms shielding using the D1-line transition should also be possible. Indeed, our shielding effect could provide an alternative explanation for the observations of Ref.~\cite{salomon2014gray}, where an atom-number enhancement is obtained when a $^{39}$K MOT is subjected to resonant D1 light. The parameters of the experiment (beams in retro-reflected configuration and total saturation parameter $s=1.5$) are compatible with the shielding requirements listed above.

\providecommand{\noopsort}[1]{}\providecommand{\singleletter}[1]{#1}%

\end{document}